# A PATTERN RECOGNITION APPROACH TO SECURE CIPHER DOCUMENTS

Saravanan Kumarasamy[1] ,Asst. Prof/CSE,Erode Sengunthar Engineering College,Thuudpathi,
Dr.T.Stephen Thangaraj[2] Professor/CSE., Erode Sengunthar Engineering College,Thuudpathi,
saravanankumarasamy@gmail.com

## 1 Abstract

Natural phenomena show that many creatures form large social groups and move in regular patterns. Previous In this paper, we first propose an efficient distributed mining algorithm to jointly identify a group of moving objects and discover their movement patterns in wireless sensor networks. Afterward, we propose a compression algorithm, called 2P2D, which exploits the obtained group movement patterns to reduce the amount of delivered data. The compression algorithm includes a sequence merge and an entropy reduction phases.  we formulate a Hit Item Replacement (HIR) problem and propose a Replace algorithm that obtains the optimal solution. Moreover, we devise three replacement rules and derive the maximum compression ratio.

Index Terms: Data compression, distributed clustering, object tracking.

## 2 INTRODUCTION

RECENT advances in location-acquisition technologies, such as global positioning systems (GPSs) and wireless sensor networks (WSNs), have fostered many novel applications like object tracking, environmental monitoring, and location-dependent service. These applications generate a large amount of location data, and thus, lead to transmission and storage challenges, especially in resourceconstrained environments like WSNs. To reduce the data volume, various algorithms have been proposed for data Compression and data aggregation [1], [2], [3], [4], [5], [6]. However, sequential patterns 1) consider the characteristics of all objects, 2) lack information
about a frequent pattern's significance regarding individual trajectories, and 3) carry no time information between consecutive items, which make them unsuitable for location prediction and similarity
 In addition, most of the above works are centralized algorithms [9], [10]. We thus define the problem of compressing the location data of a group of moving objects as the group data compression problem. Therefore, in this paper, we first introduce our distributed mining algorithm to approach the moving object clustering problem and discover group movement patterns. Our distributed mining algorithm comprises a Group Movement Pattern Mining (GMPMine) algorithm.Different from previous compression techniques that remove redundancy of data according to the regularity within the data, we devise a novel two-phase and 2D algorithm, called 2P2D, which utilizes the discovered group movement patterns shared by the transmitting node and the receiving node to compress data. Specifically, the 2P2D algorithm comprises a sequence merge and an entropy reduction phases. In the sequence merge phase, we propose a Merge algorithm to merge and compress the location data of a group of objects. In the entropy reduction phase, we formulate a Hit Item Replacement (HIR) problem to minimize the entropy of the merged data and propose a Replace algorithm to obtain the optimal solution.We formulate the HIR problem to minimize the entropy of location data and explore the Shannon's theorem to solve the HIR problem.

## 3 MINING OF GROUP MOVEMENT PATTERNS

To tackle the moving object clustering problem, we proposea distributed mining algorithm, which comprises the GMPMine algorithm. First, the GMPMine algorithm uses a PST to generate an object's significant movement patterns and computes the similarity of two objects by using simp to derive the local grouping results. The merits of simp include its accuracy and efficiency: First, simply considers the significances of each movement pattern regarding to individual objects so that it achieves better accuracy in similarity comparison. To combine multiple local grouping results into a consensus, the CE algorithm utilizes the Jaccard similarity coefficient to measure the similarity between a pair of objects, and normalized mutual information (NMI) to derive the final ensembling result. It trades off the grouping quality against the computation cost by adjusting a partition parameter. In contrast to approaches that perform clustering among the entire trajectories, the distributed algorithm discovers the group relationships in a distributed manner on sensor nodes. As a result, we can discover group movement patterns to compress the location data in the areas where objects have explicit group relationships.

Besides, the distributed design provides flexibility to take partial local grouping results into ensembling when the group relationships of moving objects in a specified subregion are interested. Also, it is especially suitable for heterogeneous tracking configurations, which helps reduce the tracking cost, e.g., instead of waking up all sensors at the same frequency, a fine-grained tracking interval is specified for partial terrain in the migration season to reduce the energy consumption. Rather than deploying the sensors in the same density, they are only highly concentrated in areas of interest to reduce deployment costs.

### 3.1 The Group Movement Pattern Mining (GMPMine) Algorithm

To provide better discrimination accuracy, we propose a new similarity measure simp to compare the similarity of two objects. For each of their significant movement patterns, the new similarity measure considers not merely two probability distributions but also two weight factors, i.e.,the significance of the pattern regarding to each PST. the negative log of the distance between two PSTs as the similarity score such that a larger value of the similarity score implies a stronger similar relationship, and vice versa.With the definition of similarity score, two objects are similar to each other if their score is above a specified similarity threshold. The GMPMine algorithm includes four steps.

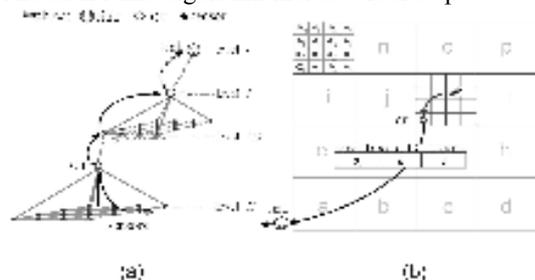

Fig. 1. (a) The hierarchical- and cluster-based network structure and the data flow of an update-based tracking network. (b) A flat view of a twolayer network structure with 16 clusters

First, we extract the movement patterns from the location sequences by learning a PST for each object. Second, our algorithm constructs an undirected, unweighted similarity graph where similar objects share an edge between each other. When the ratio of the connectivity to the size of the subgraph is higher than a threshold, the objects corresponding to the subgraph are identified as a group. We leverage the HCS cluster algorithm to partition the graph and derive the location group information.

## 4 DESIGN OF A COMPRESSION ALGORITHM WITH GROUP MOVEMENT PATTERNS

A WSN is composed of a large number of miniature sensor nodes that are deployed in a remote area for various applications, such as environmental monitoring or wildlife tracking. These sensor nodes are usually battery-powered and recharging a large number of them is difficult.On the other hand, since transmission of data is one of the most energy expensive tasks in WSNs, data compression is utilized to reduce the amount of delivered data [1], [2], [3], [4], [5], [6]. The algorithm includes the sequence merge phase and the entropy reduction phase to compress location sequences vertically and horizontally. In the sequence merge phase, we propose the Merge algorithm to compress the location sequences of a group of moving objects. The Merge algorithm avoids redundant sending of their locations, and thus, reduces the overall sequence length. It combines the sequences of a group of moving objects by 1) trimming multiple identical symbols at the same time interval into a single symbol or 2) choosing a qualified symbol to represent them when a tolerance of loss of accuracy is specified by the application. Therefore, the algorithm trims and prunes more items when the group size is larger and the group relationships are more distinct. In the entropy reduction phase, we propose the Replace algorithm that utilizes the group movement patterns as the prediction model to further compress the merged sequence.The Replace algorithm guarantees the reduction of asequence's entropy, and consequently, improves compressibility without loss of information. To reduce the entropy of a location sequence,based on which the Replace algorithm reduces the entropy efficiently. In addition, since the objects may enter and leave a sensor cluster multiple times during a batch period and a group of objects may enter and leave a cluster at slightly different times, we discuss the segmentation and alignment problems in Section 2.3. Table 1summaries the notations.

### 3.1 Sequence Merge Phase

In the application of tracking wild animals, multiple moving objects may have group relationships and share similar trajectories. In this case, transmitting their location data separately leads to redundancy. Therefore, in this section, we concentrate on the problem of compressing multiple similar sequences of a group of moving objects . Items with the same index belong to a column, and a column containing identical symbols is called an S-column; otherwise, the column is called a D-column. Finally, our algorithm generates a merged sequence containing the same information of the original sequences. In decompressing from the merged sequence, while symbol 0=0 is encountered, the items after it are output until the next 0=0 symbol. Otherwise, for each item, we repeat it n times to generate the original sequences. We regulate the accuracy by an error bound, defined as the maximal hop count between the real and reported locations of an object. To select a representative symbol for a D-column, we includes a selection criterion to minimize the average deviation between the real locations and reported locations for a group of objects at each time interval as follows Selection criterion.

3.2 Entropy Reduction Phase

In the entropy reduction phase, we propose the Replace algorithm to minimize the entropy of the merged sequence obtained in the sequence merge phase. Since data with lower entropy require fewer bits for storage and transmission , we replace some items to reduce the entropy without loss of information.In this section, we first introduce and define the HIR problem, and then, explore the properties of Shannon's entropy to solve the HIR problem. We derive three replacement rules for the HIR problem and prove that the entropy of the obtained solution is minimized.Shannon's entropy represents the optimal average code length in data compression, where the length of a symbol's codeword is proportional to its information content. A property of Shannon's entropy is that the entropy is the maximum, while all probabilities are of the same value. Consequently, 4D bits are needed to represent the location sequence. Nevertheless, since the movements of a moving object are of some regularity, the occurrence probabilities of symbols are probably skewed and the entropy is lower. Seeing that data with lower entropy require fewer bits to represent the same information, reducing the entropy thereby benefits for data compression and, by extension, storage and transmission. Motivated by the above observation, we design the Replace algorithm to reduce the entropy of a location sequence. Our algorithm imposes the hit symbols on the location sequence to increase the skewness. Specifically, the algorithm uses the group movement patterns built in both the transmitter (CH) and the receiver (sink) as the prediction model to decide whether an item of a sequence is predictable. A CH replaces the predictable items each with a hit symbol to reduce the location sequence's entropy when compressing it. After receiving the compressedsequence, the sink node decompresses it and substitutes every hit symbol with the original symbol by the identical prediction model, and no information loss occurs.A symbol is a predictable symbol once an item of the symbol is predictable. Compared with the original sequence S with entropy 3.053, the entropy of S0 is reduced to 2.854. Encoding S and S0 by the Huffman coding technique, the lengths of the output bit streams are 77 and 73 bits, respectively, i.e., 5 bits are conserved by the simple approach. However, the above simple approach does not always minimize the entropy. Consider the exampleshownin an intermediate sequence with items 1 and 19 unreplaced has lower entropy than that generated by the simple approach.For the example, the simple approach even increases the entropy.We define the above problem as the HIR problem and formulate it as follows: efinition 3 (HIR problem). The HIR problem is to find the intermediate sequence S0 such that the entropy of S0 is minimal for all possible intermediate sequences.A brute-force method to the HIR problem is to enumerate all possible intermediate sequences to find the optimal solution. However, this brute-force approach is not scalable, especially when the number of the predictable items is large. Therefore, to solve the HIR problem, we explore properties of Shannon's entropy to derive three replacement rules that our Replace algorithm leverages to obtain the optimal solution.Adding a probability with a value of zero does not change the entropy. Any permutation of the probability values does not change to the entropy  Moving all the value from one probability to another such that the former can be thought of as being eliminated decreases the entropy . If there are multiple symbols, replacing all the items of these symbols can reduce the entropy. For two probabilities, moving a value from the lower probability to the higher probability decreases the entropy For two probabilities, moving a value that is larger than the difference of the two probabilities from the higher probability to the lower one decreases the entropy According to Properties 4 and 5, we conclude that if the difference of two probabilities increases, the entropy decreases. For a probabilitydistribution  Accordingly, we derive the second replacement rule—the concentration rule: Replace all predictable items of symbol As an extension of the above properties, we also explore
the entropy variation, while predictable items of multiple symbols are replaced simultaneously. To investigate whether the converse statement exists, we conduct an experiment in a brute-force way. However, the experimental results show that even under the condition replacing predictable items of the symbols in ^s0 does not guarantee the reduction of the entropy. Therefore, we compare the difference of the entropy before and after replacing  In addition,we also prove that once replacing partial predictable items of symbols in ^s0 reduces entropy, replacing

82

all predictable items of these symbols reduces the entropy mostly since the entropy decreases monotonically

### 3.3 The Replace Algorithm

Based on the observations described in the previous section, we propose the Replace algorithm that leverages the three replacement rules to obtain the optimal solution for the HIR problem. Our algorithm examines the predictable symbols on their statistics, which include the number of items and the number of predictable items of each predictable symbol. The algorithm first replaces the qualified symbols according to the accumulation rule. Afterward, since the concentration rule and the multiple symbol rule are related to nð0:0Þ, which is increased after every replacement, the algorithm iteratively replaces the qualified symbols according to the two rules until all qualified symbols are replaced. The algorithm thereby replaces qualified symbols and reduces the entropy toward the optimum gradually. Compared with the bruteforce method that enumerates all possible intermediate sequences for the optimum in exponential complexity, the Replace algorithm that leverages the derived rules to obtain the optimal solution is more scalable and efficient. We prove that the Replace algorithm guarantees to reduce the entropy monotonically and obtains the optimal solution of the HIR problem as Theorem 1. Next, we detail the replace algorithm and demonstrate the algorithm The Replace algorithm obtains the optimal solution of the HIR problem. shows the Replace algorithm. The input includes a location sequence S and a predictor Tg, while the output, denoted by S0, is a sequence in which qualified items are replaced by 0:0. Initially, Lines 3-9 of the algorithm find the set of predictable symbols together their statistics. Then, it exams the statistics of the predictable symbols according to the three replacement rules as follows: First, according to the accumulation rule, it replaces qualified symbols in one scan of the predictable symbols as Lines 10-14. Next, the algorithm iteratively exams for the concentration and the multiple symbol rules by two loops. The first loop from Line 16 to Line 22 is for the concentration, whereas the second loop from Line 25 to Line 36 is for the multiple symbol rule. In our design, since finding a combination of predictable symbols to make hold is more costly, the algorithm is prone to replace symbols with the concentration rule.Otherwise, after an exhaustive earch for any combination of m symbols, it goes on examining the combinations of m þ 1 symbols. First, according to the accumulation rule, the predictable items are replaced After that, the statistic table is updated Second, according to the multiple symbol rule, we replace the predictable items of 0j0 and 0o0 simultaneously such that the entropy of S0 is reduced to 2.969. the predictable items of 0f0 are replaced according to the concentration rule (Lines 17-23), then the entropy of S0 is reduced to 2.893. the predictable items of symbol 0k0 are replaced according to the concentration rule. Finally, no other candidate is available, and our algorithm outputs S0 with entropy 2.854. In this example, all predictable items are replaced to minimize the entropy.

### 3.4 Segmentation, Alignment, and Packaging

In an online update approach, sensor nodes are assigned atracking task to update the sink with the location of moving objects at every tracking interval. In contrast to the online a large volume of location data for a batch period before compressing and transmitting it to the sink; and the location update process repeats from batch to batch. In real-world tracking scenarios, slight irregularities of the movements of a group of moving objects may exist in the microcosmic view. Specifically, a group of objects may enter a sensor cluster at slightly different times and stay in a sensor cluster for slightly different periods, which lead to the alignment problem among the location sequences. Moreover, since the trajectories of moving objects may span multiple sensor clusters, and the objects may enter and leave a cluster multiple times during a batch period, a location sequence may comprise multiple segments, each of which is a trajectory that is continuous in time domain. To deal with the alignment and segmentation problems, we partition location sequences into segments, and then, compress and package them into one update packet. Consider a group of three sequences shown in Fig. 12a, the segments $E_1$, $E_2$, and $E_3$ are aligned and named G-segments, whereas segments A, B, C, and D are named S-segments. Figs. 12b, 12c, and 12d show an illustrative example to construct the frame for the three sequences. First, the Merge algorithm combines $E_1$, $E_2$, and $E_3$ to generate an intermediate sequence S00 E. Next, S00 E together with A, B, C, and D is viewed as a sequence and processed by the Replacealgorithm to generate an intermediate sequence S0, whichcomprises S0A , S0B, S0C , S0D , and S0E. Finally, intermediate sequence S0 is compressed and packed.For a batch period of D tracking intervals, the location data of a group of n objects are aggregated in one packet packet headers are eliminated. The payload may comprise multiple G-segments or S-segments, each of which includes a beginning time stamp (a bits), a sequence of consequent locations (b bits for each), an object or group ID ( c bits), and a field representing the length of a segment (l bits). By exploiting the correlations in the location data, we can further compress the location data and reduce the amount of data and H denotes the data size of the packet header. As for the online update approach, when a sensor node detects an object of interest, it sends an update packet upward to the sink. The payload of a packet includes time stamp, location, and object . Some approaches, such as [55], employ techniques like location prediction to reduce the number of transmitted update packets. For D tracking intervals, the amount of data for tracking n objects.

Therefore, the group size, the number of segments, and the compress ratio are important factors that influence the performance of the batch-based approach. In the next section, we conduct experiments to evaluate the performance of our design.

## 4 EXPERIMENT AND ANALYSIS

We implement an event-driven simulator evaluate the performance of our design. To the best of ur knowledge, no research work has been dedicated to discovering application-level semantic for location data compression. We compare our batch-based approach with an online approach for the overall system performance evaluation and study the impact of the group size (n), as well as the group dispersion radius (GDR ), the batch period (D), and the error bound of accuracy (eb). We also compare our Replace algorithm with Huffman encoding technique to show its effectiveness. Since there is no related work that finds real location data of group moving objects, we generate the location data, i.e., the coordinates (x; y), with the Reference Point Group Mobility Model  for a group of objects moving in a two-layer tracking network with 256 nodes. A location-dependent mobility model is used to simulate the roaming behavior of a group leader; the other member objects are followers that are uniformly distributed within a specified group dispersion radius (GDR) of the leader, where the GDR is the maximal hop count between followers and the leader. We utilize the GDR to control the dispersion degree of the objects. Smaller GDR impliesstronger group relationships, i.e., objects are closer together. The speed of each object is 1 node per time unit, and the tracking interval is 0.5 time unit. In addition, the starting point and the furthest point reached by the leader object are randomly selected, and the movement range of a group of objects is the euclidean distance between the two points. The data sizes of object (or group) ID, location ID, time stamp, and packet header are 1,1, 1, and 4 bytes, respectively. Moreover, we use the amount of data in kilobyte (KB) and compression ratio (r) asthe evaluation metric, where the compression ratio is defined as the ratio between the uncompressed data size and the compressed data size. First, we compare the amount of data of our batch-based approach (batch) with that of an online update approach (online). In addition, some approaches, such as [55], employ techniques like location prediction to reduce the number of transmitted update packets. We use the discovered movement patterns as the prediction model for prediction in the online update approach.our batch-based approach outperforms the online approach with and without prediction. The amount of data of our batchbased approach is relatively low and stable as the GDR increases. Compared with the online approach, the compressionratios of our batch approach and the online approach with prediction are about 15.0 and 2.5 as GDR ¼ 1.Next, our compression algorithm utilizes the group relationships to reduce the data size. Fig. 13b shows the impact of the group size. The amount of data per object decreases as the group size increases. Compared with carrying the location data for a single object by an individual packet, our batch-based approach aggregates and compresses packets of multiple objects such that the amount of data decreases as the group size increases. Moreover, our algorithm achieves high compression ratio in two ways. First, while more sequences that are similar or sequences that are more similar are compressed simultaneously, the Merge algorithm achieves higher compression ratio. Second, with the regularity in the movements of a group of objects, the Replace algorithm minimizes the entropy which also leads to higher compression ratio. Note that we use the GDR to control the group dispersion range of the input workload. The leader object's movement path together with the GDR sets up a spacious area where the member objects are randomly distributed. Therefore, a larger GDR implies that the location sequences have higher entropy, which degrades both the prediction hit rate and the compression ratio. Therefore, larger group size and smaller GDR result in higher compression ratio. Fig. 14a shows the impact of the batch period (D). The amount of data decreases as the batch period increases. Since more packets are aggregated and more data are compressed for a longer batch period, our batch-based approach reduces both the data volume of packet headers and the location data. Since the accuracy of sensor networks is inherently limited, allowing approximation of sensors' readings or tolerating a loss of accuracy is a compromise between data accuracy and energy conservation. We study the impact of accuracy on the amount of data. As GDR varies from 0.1 to 1 the compression ratios of theHuffman encoding with and without our Replace algorithm; while  the prediction hit rate. Compared with Huffman, our Replace algorithm achieves higher compression ratio, e.g., the compression ratio of our approach is about 4, while that of Huffman is about 2.65 we show that the compression ratio that the Replace algorithm achieves reduces as the prediction hit rate. As the prediction hit rate is about 0.6, the compression ratio of our design is about 2.7 that is higher than 2.3 of Huffman.

## 5 BLOWFISH ALGORITHM

The data transformation process for PocketBrief uses the *Blowfish Algorithm* forEncryption and Decryption, respectively. *Blowfish* is a symmetric block cipher that can be effectively used for encryption and safeguarding of data. It takes a variable-length key, from 32 bits to 448 bits, making it ideal for securing data. *Blowfish* was designed in 1993 by Bruce Schneier as a fast, free alternative to existing encryption algorithms. *Blowfish* is unpatented and license-free, and is available free for all uses.

*Blowfish* has 16 rounds.

The input is a 64-bit data element, x.

Divide x into two 32-bit halves: xL, xR.

Then, for i = 1 to 16:

xL = xL XOR Pi

xR = F(xL) XOR xR

Swap xL and xR

After the sixteenth round, swap xL and xR again to undo the last swap.

Then, xR = xR XOR P17 and xL = xL XOR P18.

Finally, recombine xL and xR to get the ciphertext.

PocketBrief

5 of 7 Decryption is exactly the same as encryption, except that P1, P2,..., P18 are used in the reverse order. Implementations of *Blowfish* that require the fastest speeds should unroll the loop and ensure that all subkeys are stored in cache.

## 6 CONCLUSIONS

In this work, we exploit the characteristics of group movements to discover the information about groups of moving objects in tracking applications. We propose a distributed mining algorithm, which consists of a local GMPMine algorithm and a CE algorithm, to discover group movement patterns. With the discovered information, we devise the 2P2D algorithm, which comprises a sequence merge phase and an entropy reduction phase. In the sequence merge phase, we propose the Merge algorithm to merge the location sequences of a group of moving objects with the goal of reducing the overall sequence length. In the entropy reduction phase, we formulate the HIR problem and propose a Replace algorithm to tackle the HIR problem. In addition, we devise and prove three replacement rules, with which the Replace algorithm obtains the optimal solution of HIR efficiently. Our experimental results show that the proposed compression algorithm effectively reduces the amount of delivered data and enhances compressibility and, by extension, reduces the energy consumption expense for data transmission in WSNs.